\documentclass{article}

\usepackage{PRIMEarxiv}

\usepackage{underscore}
\usepackage{amsmath,amsfonts}
\usepackage{array}
\usepackage{textcomp}
\usepackage{stfloats}
\usepackage{url}
\usepackage{verbatim}
\usepackage{graphicx}
\usepackage{cite}
\usepackage{lscape}
\usepackage{caption}
\usepackage{subfig}
\usepackage{multirow}
\usepackage{underscore}
\usepackage{hhline}
\usepackage{adjustbox}

\graphicspath{ {./images/}} 

\makeatletter
\DeclareRobustCommand{\cev}[1]{%
  {\mathpalette\do@cev{#1}}%
}
\newcommand{\do@cev}[2]{%
  \vbox{\offinterlineskip
    \sbox\z@{$\m@th#1 x$}%
    \ialign{##\cr
      \hidewidth\reflectbox{$\m@th#1\vec{}\mkern4mu$}\hidewidth\cr
      \noalign{\kern-\ht\z@}
      $\m@th#1#2$\cr
    }%
  }%
}
\makeatother

\title{Local Temporal Feature Enhanced Transformer with ROI-rank Based Masking for Diagnosis of ADHD}

\begin{document}

\newcommand*{\affaddr}[1]{#1} 
\newcommand*{\affmark}[1][*]{\textsuperscript{#1}}
\newcommand*{\email}[1]{\texttt{#1}}

\author{
Byunggun Kim\affmark[1] and Younghun Kwon\affmark[1,2]\\
\affaddr{\affmark[1]Department of Applied Artificial Intelligence}\\
\affaddr{\affmark[2]Department of Applied Physics}\\
\email{\{byunggunkim, yyhkwon\}@hanyang.ac.kr}\\
\affaddr{Hanyang University, Ansan, Kyunggi-Do, 425-791, Republic of Korea}%
}

\maketitle

\begin{abstract}
\label{sec:abstract}
In modern society, Attention-Deficit/Hyperactivity Disorder (ADHD) is one of the common mental diseases discovered not only in children but also in adults.
In this context, we propose a ADHD diagnosis transformer model that can effectively simultaneously find important brain spatiotemporal biomarkers from resting-state functional magnetic resonance (rs-fMRI).
This model not only learns spatiotemporal individual features but also learns the correlation with full attention structures specialized in ADHD diagnosis.
In particular, it focuses on learning local blood oxygenation level dependent (BOLD) signals and distinguishing important regions of interest (ROI) in the brain.
Specifically, the three proposed methods for ADHD diagnosis transformer are as follows.
First, we design a CNN-based embedding block to obtain more expressive embedding features in brain region attention.
It is reconstructed based on the previously CNN-based ADHD diagnosis models for the transformer.
Next, for individual spatiotemporal feature attention, we change the attention method to local temporal attention and ROI-rank based masking.
For the temporal features of fMRI, the local temporal attention enables to learn local BOLD signal features with only simple window masking.
For the spatial feature of fMRI, ROI-rank based masking can distinguish ROIs with high correlation in ROI relationships based on attention scores, thereby providing a more specific biomarker for ADHD diagnosis.
The experiment was conducted with various types of transformer models.
To evaluate these models, we collected the data from 939 individuals from all sites provided by the ADHD-200 competition.
Through this, the spatiotemporal enhanced transformer for ADHD diagnosis outperforms the performance of other different types of transformer variants. (77.78ACC 76.60SPE 79.22SEN 79.30AUC)
\end{abstract}

\section{Introduction}
\label{Introduction}
\label{sec:introduction}
Attention- deficit/hyperactivity disorder (ADHD) is a common mental disease that occurs from children to adults \cite{edition1980diagnostic,faraone2000attention,sayal2018adhd}. 
It causes difficulties in many ways among daily life.
In the recent, the number of patients diagnosed with ADHD has been increasing \cite{polanczyk2007worldwide,chung2019trends}.
However, the exact cause of ADHD has not yet been discovered, either clinically or psychologically \cite{thapar2012causes}.

Recently, the use of Functional Magnetic Resistance Imaging (fMRI) data as a new approach to finding biomarkers of ADHD has attracted attention \cite{greicius2008resting,barkhof2014resting,lin2014global,posner2014connecting,canario2021review}.
This is because fMRI contain biological activation information according to the location of the brain and has shown meaningful results in studies of various neuropsychiatric diseases \cite{greicius2008resting,barkhof2014resting}.
It has become possible to find reliable biomarkers of neuropsychiatric diseases by using the Resting-State Functional Connectivity (RSFC) that analyzes regional correlations in brain \cite{RN153}.
However, traditional RSFC analysis uses only a small amount of data obtained from a specific site, making it difficult to detect heterogeneity across individual studies \cite{castellanos2008cingulate,castellanos2012large}.

\begin{figure}[!t]
\centering
\includegraphics[width=0.8\textwidth]{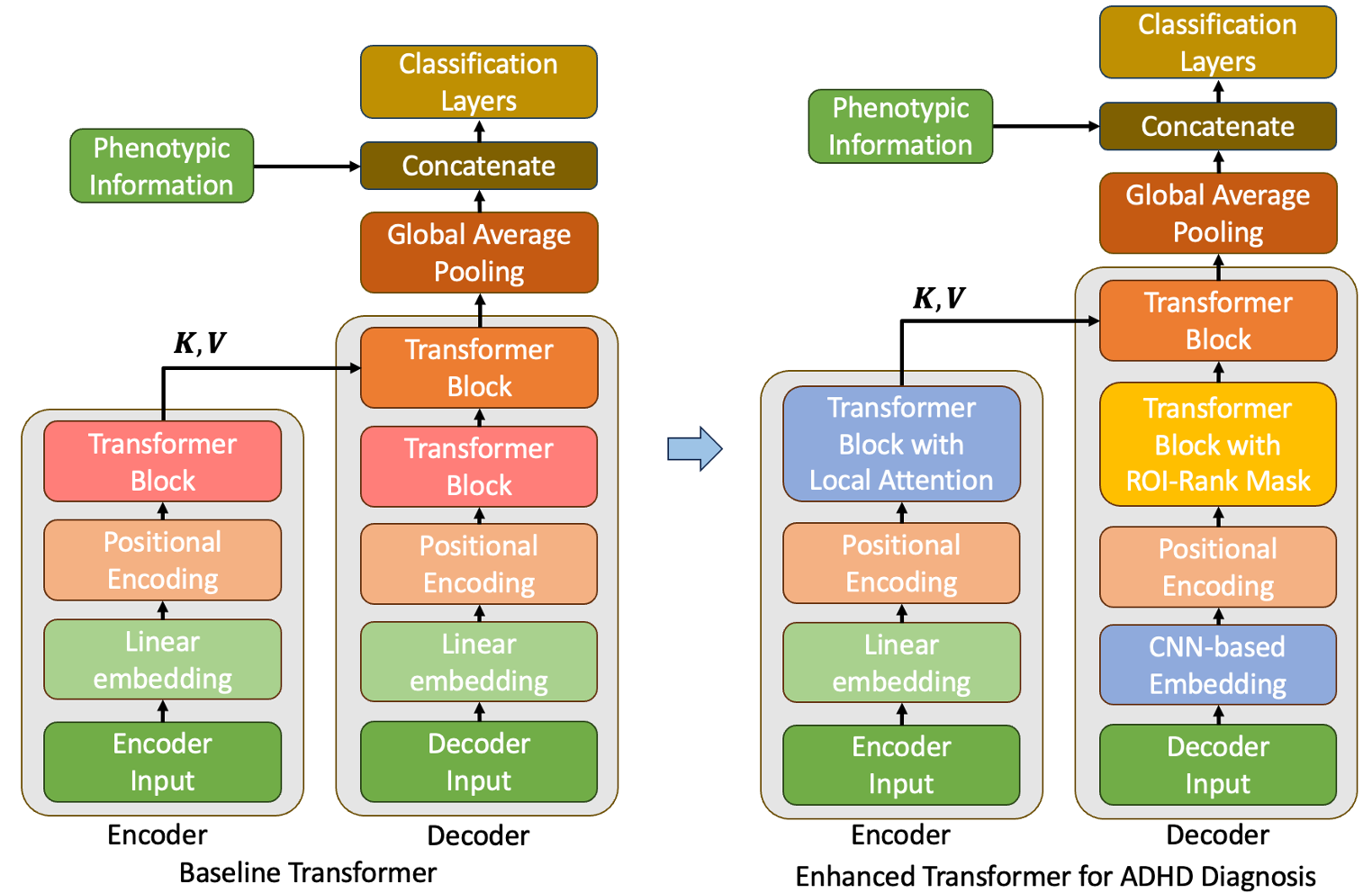}
\caption{The enhanced ADHD diagnosis transformer model architecture. It modified three modules (CNN-based embedding block, local temporal attention, and ROI-rank based masking) from baseline transformer.}
\label{img:architecture}
\end{figure}

Therefore, most of studies \cite{eloyan2012automated,loh2022automated} are a shift toward using a large number of rs-fMRI data samples collected from various sites to find more general biomarkers.
Especially, deep learning models have shown great potential in discovering new biomarkers for ADHD \cite{koppe2021deep,riaz2017fcnet,riaz2020deepfmri,RN8}.
Subsequently, the development of an ADHD diagnostic model using rs-fMRI has led to new understanding.

Specifically, deep learning-based ADHD diagnosis models have proposed the architectures for extracting temporal or spatial features from rs-fMRI \cite{riaz2017fcnet,RN1,riaz2020deepfmri,RN8,RN7,liu2022attention}.
These usually consist of Convolutional Neural Network (CNN) and Recurrent Neural Network (RNN) to extract brain functions contained in the Blood Oxygen Level Dependent (BOLD) signal.
However, due to structural limitations, they had difficulty to train spatiotemporally global features from rs-fMRI data.
Accordingly, the attention mechanism \cite{vaswani2017attention}, which can learn important features through relationships, is being actively used in learning rs-fMRI data \cite{ji2022fc,qiang2022novel,qin2022ensemble,liu2023spatial}.
Among them, models that modify the transformer structure to make full use of the attention mechanism for spatiotemporal feature extraction of rs-fMRI are increasing. 
However, since rs-fMRI data has a four-dimension $(x, y, z, t)$, the attention mechanism optimized only for learning the partial relation of rs-fMRI has been proposed.

To overcome these limitations, we constructed an encoder-decoder transformer model capable of learning the relationship between the full-range of rs-fMRI.
This enables to learn independent relationship within spatial or temporal features.
Also, it can attend to relation between spatial and temporal features. 
In addition, we proposed three modifications of the transformer structure suitable for ADHD diagnosis. 
These are one embedding method constructed based on 1D-CNN and two attention methods specialized for each feature: local temporal attention for temporal feature and ROI-rank based masking for spatial feature.

Specifically, local temporal attention is an attention that uses window masking to restrict only the relationships between short term BOLD signals.
This show that heavily focusing on relationship within local BOLD signals is important in first stage to learn temporal features.
And BOLD signal learning is more effective by hierarchically applying full temporal attention after local temporal attention.
Next, ROI-rank based masking is an attention method that effectively select a small number of ROIs based on attention scores.
Therefore, it learns only the relationships between ROIs that more relevant to ADHD differentiation. 
This provides clues to support the assumption that patients with ADHD may have more sparsely strong functional connectivity in brain than healthy control (HC).

In summary, the overall structure is shown in figure \ref{img:architecture}.
This new ADHD diagnosis transformer showed notable biomarker simultaneously not only in BOLD signal but also in ROI that more important than other.
To prove this new ADHD diagnosis transformer's performance, we collect data from multiple sites provided by the ADHD-200 competition.
Also, we evaluate the various ablation type of models to prove that our proposed methods can more effectively diagnosis the ADHD.
Through spatiotemporally enhanced transformer for diagnose ADHD, we achieved best diagnostic performance (77.78ACC 76.60SPE 79.22SEN 79.30AUC).

The contents of each section of the paper are as follows.
In Section \ref{sec:related works}, we investigated the studies to find biomarkers of ADHD and the development of a deep learning-based ADHD diagnostic model.
In section \ref{sec:preprosed method}, we explain that our proposed the transformer model and modified methods for ADHD diagnosis.
In section \ref{sec:experimental setting and results}, we describe the specific experiments using the ADHD200 corpus and the analysis.
Finally, in section \ref{sec:conclusion}, we conclude this study.

\section{Related Works}
\label{sec:related works}

\subsection{Traditional Study with rs-fMRI for ADHD}
\label{sec:related works 1}

It is difficult to find the clear evidence of ADHD.
However, with the advent of the fMRI, which can indirectly obtain brain activity information from patients in a noninvasive manner, many of studies have shown meaningful results to find biomarkers of ADHD.
Especially, with rs-fMRI, Default Mode Network (DMN), a network analysis that focus on the resting state brain activity, contributed to find the important brain regions that distinguish between focused and non-focused states of the brain \cite{raichle2001default,fox2005human,raichle2015brain}. 
These regions are also closely related to mental disorders that exhibit impaired concentration, such as ADHD \cite{castellanos2008cingulate,sun2012abnormal,marcos2018local,sutcubasi2020resting}.

For example, Castellanos et al., \cite{castellanos2008cingulate} performed a seed-based rs-fMRI analysis between the DMN and three specific frontal regions within the Cognitive Control Network (CCN).
Through this, they found that in HC, regions of the cognitive control network had anticorrelations with some regions contained in DMN. 
On the contrary, in ADHD patients, the anticorrelation between regions of these two networks was revealed to be weak.
Specifically, the relationship between the dorsal anterior cingulate cortex (dACC) in the CCN and the DMN did not show a significant difference in ADHD patients compared to HC.

However, there are some limitations in these traditional studies \cite{smith2012future,specht2020current}.
One reason is that because the collected data that are used in studies are inconsistency, it led to differences.
Therefore, it is hard to overcome the heterogeneity results.
Another reason is that the data sample used for analysis is too small.
This is because it is difficult to recruit ADHD patients who meet the experimental conditions.
To overcome these limitations, recent studies were changed to used open datasets.

\subsection{Diagnosis ADHD with Deep Learning Model}
\label{sec:related works 2}

Deep learning-based ADHD diagnostic models provide reasonable results that support the existing hypothesis on the cause of ADHD.
Moreover, the available rs-fMRI data is increasing, the limitations of analysis using only a small amount of data within the site are being overcome.
So, it is shown more general biomarkers related with ADHD. 
Also, as well-suited deep learning model structures are proposed, the accuracy of ADHD diagnosis is gradually increasing.

Early deep learning-based ADHD diagnosis models are purposed to find connection networks between brain regions through learning \cite{riaz2017fcnet,riaz2020deepfmri,zhang2020separated,ariyarathne2020adhd}.
For this purpose, 1-D CNN which effectively learned new temporal features from brain region-specific signals of rs-fMRI was widely used.
However, since they mainly used seed-based fMRI features, they showed different connection networks depending on how the brain regions were divided.
Therefore, without considering the functional connectivity, raw rs-fMRI data was used to eliminate template-specific bias.
With 3-D CNN \cite{RN1,wang2019dilated}, spatial feature on the brain volume can be extracted from raw rs-fMRI.
However, compared to seed-based fMRI features, raw rs-fMRI has a considerably large data size, it is hard to train spatial and temporal features together.

Another approach to design an ADHD diagnosis model is to learn temporally dynamic features of rs-fMRI signals.
To do so, the RNN structure is utilized to learn changes in the brain network over time. 
Mao et al., \cite{RN8} proposed a model by combining 3D CNN and LSTM to learn spatiotemporal features of fMRI. 
However, it has a disadvantage that it is limited to the spatial features of a single frame of fMRI. 
So, they also proposed a 4D CNN and showed that it is also effective for learning the full range of features of fMRI. 
Otherwise, Liu et al., \cite{liu2020multi} proposed a nested residual convolutional denoising autoencoder (NRCDAE) to help extract spatial features of a single frame of fMRI.
And then, they used a convolutional GRU to efficiently train the temporal features.

Recently, as the attention mechanism-based transformer model has been successful in various fields \cite{vaswani2017attention,dosovitskiy2020image,peebles2023scalable,gulati2020conformer}, efforts to break away from CNN and RNN structures in ADHD diagnosis have been increasing.
Because of the relation-based learning, attention mechanism has the advantage of being able to interpret of ADHD diagnosis \cite{ji2022fc,qiang2022novel,qin2022ensemble,liu2023spatial}.
Through this, it is possible to confirm which time or region are heavily affects the ADHD diagnosis in rs-fMRI.
However, because fMRI is a 4D data structure, it is also difficult to train relational information that simultaneously consider spatiotemporal features.
For that reason, we proposed an encoder-decoder type transformer model that not only extract spatial or temporal features independently, but also learn the joint relation between spatiotemporal features.

\section{Proposed Method}
\label{sec:preprosed method}
Transformer has proved the power of learning with attention mechanism.
In this situation, we developed a suitable transformer model for ADHD diagnosis.
First, we design fully attention transformer for rs-fMRI data.
And then, we also modified the three parts (Embedding, Encoder, Decoder) in transformer to more effective diagnosis for the ADHD.
Through this, this new transformer model can achieve hightest diagnosis accuracy in ADHD.

\subsection{Attention Mechanism for Diagnosis of ADHD with rs-fMRI}
\label{sec:preprosed method 1}

\begin{figure}[!t]
\centering
\includegraphics[width=0.8\textwidth]{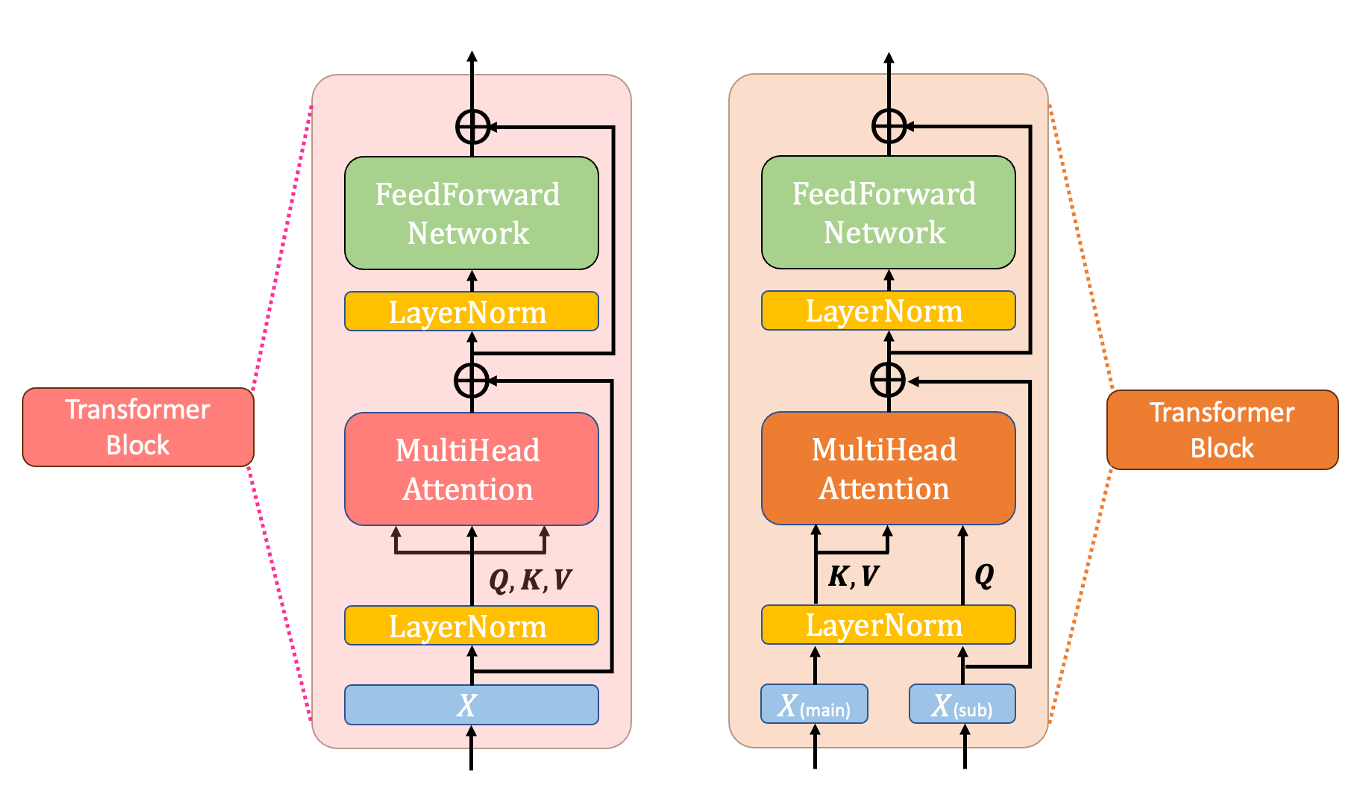}
\caption{The structure of the feature-independent self-attention (left) and spatiotemporal co-attention (right). The feature-independent self-attention (left), same as original self-attention, can focus on in each perspective feature learning. Whereas the spatiotemporal co-attention (right) learns how to relate between spatial and temporal features in rs-fMRI.}
\label{img:twotransformerblock}
\end{figure}

As 4D $(x, y, z, t)$ shape data, rs-fMRI is regional brain activation information.
Therefore, it is important to learn which time or which brain region is important through individual learning of temporal $(t)$ and spatial features $(x, y, z)$ for ADHD diagnosis.
In addition, comprehensive learning that considers both features are also necessary for learning the overall context of the data.
Therefore, we proposed an attention mechanism suitable for fMRI data for ADHD diagnosis.

\begin{equation}
\label{attention}
\mathrm{Attention} (Q, K, V) = \mathrm{Softmax}( {{QK^T} \over {\sqrt{n}}} )V
\end{equation}

In equation \ref{attention}, the basic attention mechanism \cite{vaswani2017attention} learns the relationship in the input data through three elements: Query $(Q)$, Key $(K)$, and Value $(V)$ to find important features.
Specifically, it is as follows.
Query means what you want to find with the question.
Key helps you find the information that best matches the question.
Finally, Value is the information that valuable to obtain and is paired with Key.
Depending on how to prepare these three elements of attention, fMRI data can have two types of attention with two perspectives.

To design the attention mechanism for diagnosis ADHD, fMRI data can be embedded to a temporal perspective feature or a spatial perspective feature denoted as $X^{(T)} \in \mathbb{R}^{(T \times h_{a} )}$ and, $X^{(S)} \in \mathbb{R}^{(S \times h_{a} )}$ respectively.
Here, $T$ is the length of the fMRI data signal, $S$ is the number of regions of interest (ROI), and $h_a$ represents the size of the attention embedding vector.

Then, with these two feature perspectives, we can construct feature-independent self-attention or spatiotemporal co-attention.
First, feature-independent self-attention learns perspective each specific feature by using the self-attention.
Specifically, self-attention in the temporal perspective feature $(X^{(T)})$ learns which time plays an important role in the fMRI signal for ADHD diagnosis.
And, self-attention in the spatial perspective feature $(X^{(S)})$ learns which ROI is important for ADHD diagnosis.
Currently, the $Q, K, V$ representations for feature-independent self-attention are as follow equation \ref{temporalqkv} and \ref{spatialqkv}.

\begin{equation}
\label{temporalqkv}
 X^{(T)} \to Q^{(T)}, K^{(T)}, V^{(T)}
\end{equation}

\begin{equation}
\label{spatialqkv}
 X^{(S)} \to Q^{(S)}, K^{(S)}, V^{(S)}
\end{equation}

However, feature-independent self-attention alone can not extract the overall fMRI’s context that synthesize the two perspectives.
Therefore, to compensate this shortcoming, spatiotemporal co-attention, which is attention that considers the relationship between both temporal perspective feature $(X^{(T)})$ and spatial perspective feature $(X^{(S)})$ is needed.
Spatiotemporal co-attention connects the relationships between features of different perspectives by making the features corresponding to the three elements of $Q$, $K$, $V$ differently.
Thereby, it enable to learn more globally.

\begin{equation}
\label{spatial to temporal}
 X^{(T)}_{(\mathrm{main})} \to (K^{(T)}, V^{(T)}), X^{(S)}_{\mathrm{(sub)}} \to Q^{(S)}
\end{equation}

\begin{equation}
\label{temporal to spatial}
 X^{(S)}_{(\mathrm{main})} \to (K^{(S)}, V^{(S)}), X^{(T)}_{\mathrm{(sub)}} \to Q^{(T)}
\end{equation}

The equation \ref{spatial to temporal}, \ref{temporal to spatial} shows how each feature corresponds to the attention elements.
Specifically, the main feature corresponds to $K$ and $V$ and is used as a core feature for ADHD diagnosis.
And $Q$ corresponds to the sub feature and is used as a clue to make a reasonable choice about which part of the main feature in the two-viewpoint data
Therfore, with rs-fMRI data, two types of spatiotemporal co-attention are possible.

\subsection{Encoder-Decoder Transformer for The ADHD Diagnosis}
\label{sec:preprosed method 2}

\begin{figure}[!t]
\centering
\includegraphics[width=0.7\textwidth]{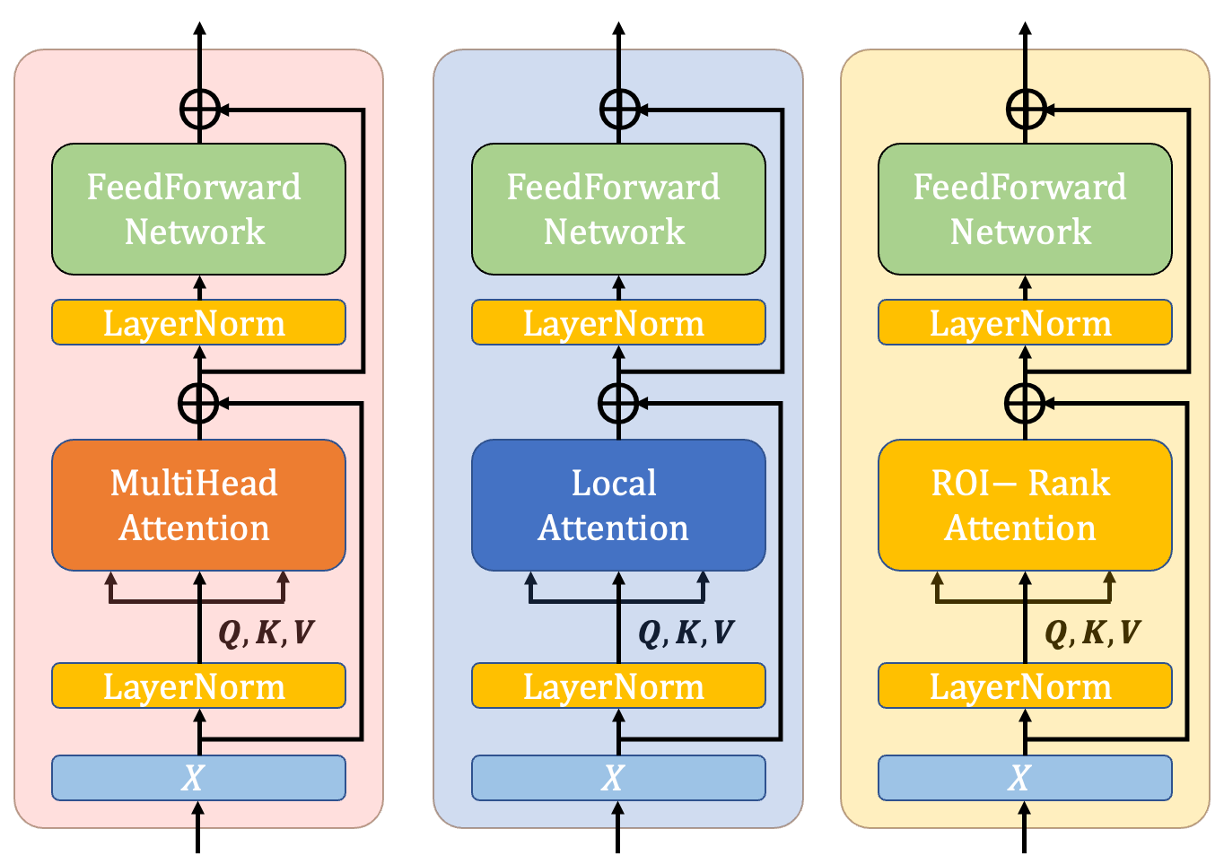}
\caption{The three different type of transformer block. The (left) is the original transformer block. The (middle) is the local temporal attention-based transformer block. The (right) is the ROI-Rank masking attention-based transformer block.}
\label{img:threeattention}
\end{figure}

With these two type of attention methods, we constructed an encoder-decoder transformer for the ADHD diagnosis.
The left figure in figure \ref{img:architecture} shows overall architecture of the encoder-decoder transformer.
Specifically, both the encoder and decoder consist of a linear embedding layer, positional encoding, and a transformer block.
First, the linear embedding layer enables learning of fMRI data as a vector representation of feature views.
Then, in second, positional encoding helps distinguish each brain region and time step.
For this, we used the sinusoidal method used in the transformer model.

Next, the transformer block is composed of a combination of the attention layer and Feed-Forward Network (FFN), as shown in the figure.
The transformer block is divided into two depending on the attention method.
In figure \ref{img:twotransformerblock}, the pink block used the feature-independent self-attention at located in front of the encoder and decoder.
Then, the orange block used spatiotemporal co-attention at the last stage of the decoder.
The output features of the encoder are the main input of attention $(X_{\mathrm{(main)}})$ and the output features of the decoder are the sub input of attention $(X_{\mathrm{(sub)}})$.
Therfore it can synthesizes all the features of the fMRI.
Additionally, the two transformer blocks have the same pre-LN structure, which applies Layer Normalization (LN) right before the attention block and FFN input for the stable.

And, the last part of the model is a process to classify the ADHD.
First, with Global Average Pooling (GAP), the output features of the final decoder are synthesized. 
Then second, phenotypic information is added to provide general personal information. 
Finally, combined features passed through three dense layers.
The specific model settings used in the experiment can be found in Table \ref{tab:modelparameter}.

Through experiments, we confirmed that the best diagnosis performance is possible when temporal perspective feature is used as input to the encoder and spatial perspective feature is used as input to the decoder.
Therefore, based on this result, we improved this architecture to be more suitable for ADHD diagnosis.
The right figure in figure \ref{img:architecture} and figure \ref{img:threeattention} show the modified transformer and the detail in the attention method inside the transformer block.
This changes shows that the clue how to diagnose the ADHD with deep learning method.

Specifically, we modified the three parts in transformer.
The first change is the spatial feature embedding block.
It is a CNN-based module used for embedding spatial perspective feature.
The second change is local temporal attention.
With simple window mask, it can attend only in the local BOLD signal.
Finally, the third change is ROI-rank based masking attention.
This not only allows for selecting important ROIs based on attention scores, but also confirms that ROI selection is very important for ADHD diagnosis.
The description of specific changes can be found in following subsections.

\subsection{CNN-based Embedding Block}
\label{sec:preprosed method 3}

\begin{figure}[!t]
\centering
\includegraphics[width=0.5\textwidth]{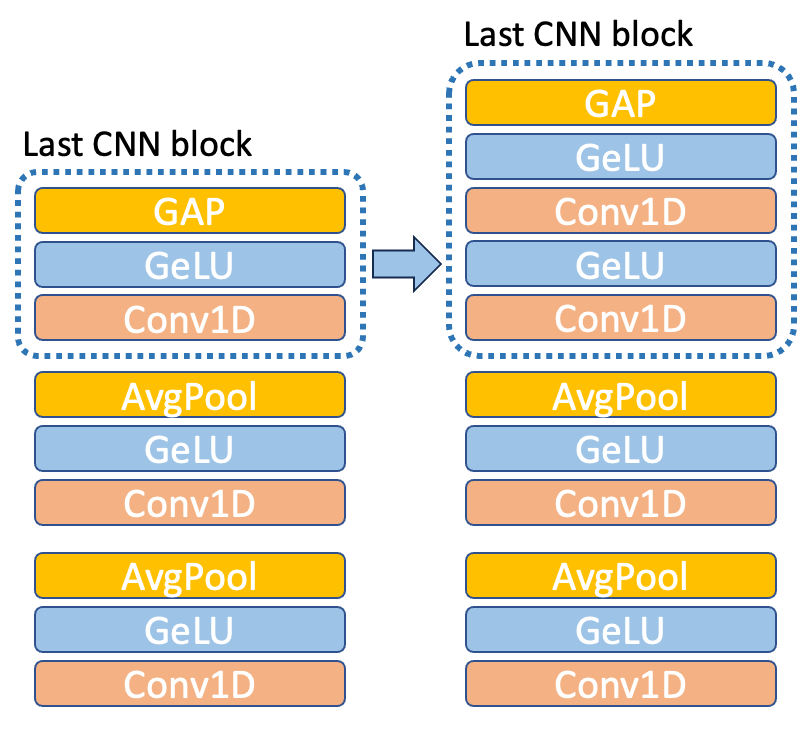}
\caption{The detail of CNN-based Embedding blocks. We change the last CNN block with 2 convolutional layers with GeLU activation.}
\label{img:cnnembedding}
\end{figure}

The original spatial feature embedding layer in baseline transformer embeds the entire fMRI signal regardless of time variation.
Therefore, to extract temporal feature more accurately way, we design the embedding block with CNN that can efficiently train the local temporal information.
With this CNN-based embedding block, we can enhance the representation of spatial perspective feature $(X^{(S)})$, thereby improving the ADHD diagnosis performance.

To find the well-suited CNN-based embedding block for this model, we referenced previously proposed CNN models.
A proposed CNN-based embedding block is as shown in figure \ref{img:cnnembedding}.
First, in the left figure in figure \ref{img:cnnembedding} is the original CNN-based embedding block.
Specifically, the CNN block is composed of Conv1D – GeLU – Average Pooling in the order of Conv1D – GeLU – Average Pooling, and in the case of the last CNN block, average pooling is changed to GAP to use the channel feature as the embedding feature.
Another structure in right figure in figure \ref{img:cnnembedding} is a improved CNN-based Embedding block with reference to the existing CNN-based ADHD diagnosis model.
As can be seen in blue box, it is a two-layer CNN without pooling for the last CNN block.
With this, it can focus on extracting high- dimensional channel features.

Through experiments, we were able to achieve more improved ADHD diagnosis performance by using CNN-based embedding blocks. 
In particular, improved CNN-based embedding block showed a much more effective than before.
This shows that enhancing the expression of high-dimensional channel features in CNN-based embedding block is important to get more finer fMRI signal embedding.

\subsection{Local Temporal Attention}
\label{sec:preprosed method 4}

\begin{figure}[!t]
\centering
\includegraphics[width=0.9\textwidth]{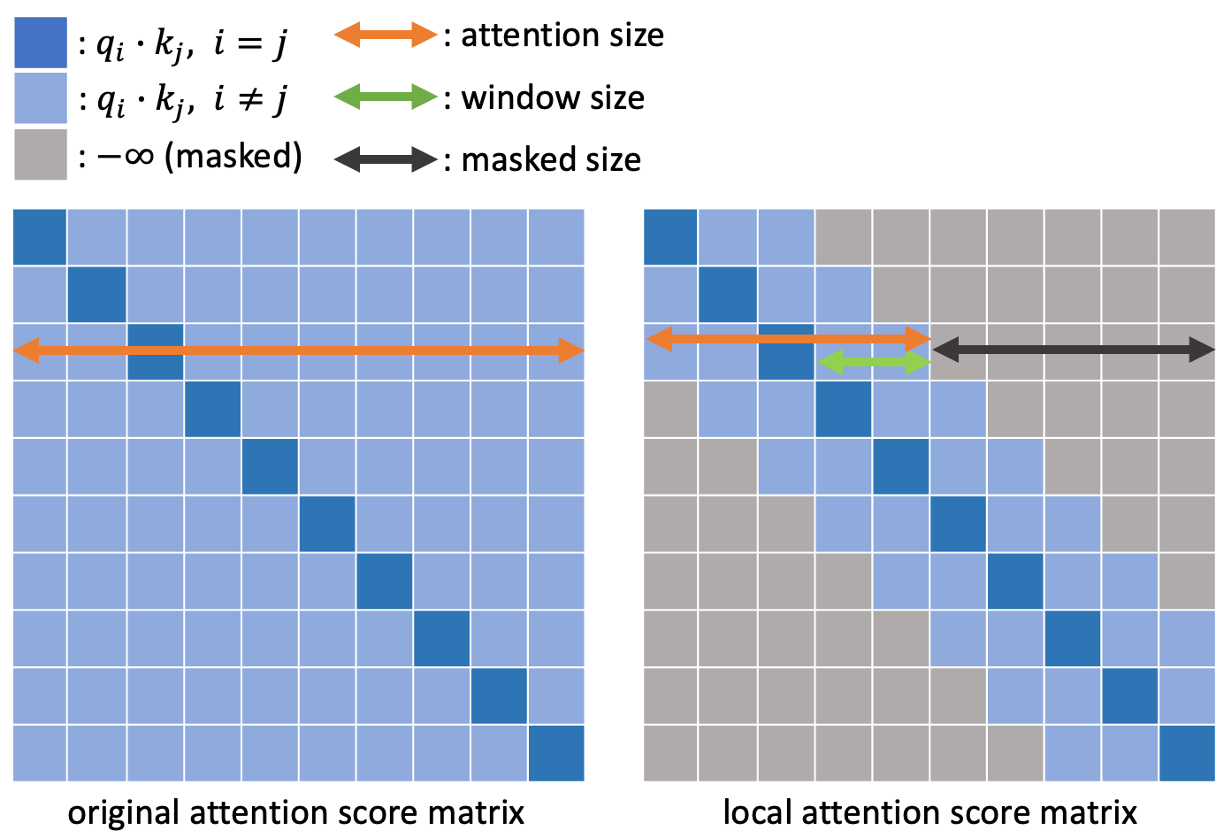}
\caption{The temporal local attention’s masking method in score matrix. Specifically, the only blue grid’s query-key scores that neighbor with the diagonal element is used to attend. And the other gray grid’s query-key scores are ignored.}
\label{img:localtemporalattention}
\end{figure}

In the case of the attention model, effective attention learning is possible if an appropriate attention pattern can be found with pre-knowledge \cite{child2019generating}.
Therefore, we want to design an prior attention pattern for fMRI.
Basically, fMRI is based on the BOLD signal, which is a pulse with a specific shape in a short period of time in the brain.
For that reason, we used a local window pattern to learn relationships within a certain range on the BOLD signal.
The equation \ref{eq:window1} to \ref{eq:window4} and figure \ref{img:localtemporalattention} show the pattern shape and application method.

\begin{equation}
\label{eq:window1}
\mathrm{Attention} (X^{(T)}, M) = (a(x_i^{(T)}, M_i))
\end{equation}

\begin{equation}
\label{eq:window2}
 a(x_i^{(T)}, M_i) = \mathrm{Softmax}( {q_i^{(T)} K_{M_i}^{(T)T}\over{\sqrt{n}}}) V_{M_i}^{(T)}
\end{equation}

\begin{equation}
\label{eq:window3}
 q_i^{(T)} = W_q x_i
\end{equation}

\begin{equation}
\label{eq:window4}
 K_{M_i}^{(T)}=(W_k x_j^{(T)})_{j \in M_i}, V_{M_i}^{(T)}=(W_v x_j^{(T)})_{j \in M_i}
\end{equation}

In the formula, $M_i= \{p,p_1,...,i,...,q-1,q\}$ $\mathrm{for}$ $p =\max (1,i-l), q=\min(T,i+1) $.
This mask restricts to learn only the relationship between the features of a certain time step (i) of an fMRI signal and the target temporal range.
Specifically, it is as follows.
First, with each weight parameters $(W_{q}, W_{k}, W_{v})$, the $i \in \{1, ...,T\}$ th temporal perspective feature $(X_i^{(T)})$ is represented to querise, keys and values $(q_i^{(T)}, k_{M_i}^{(T)}, v_{M_i}^{(T)})$ that located in the window size $(l)$ from time step $(i)$.
And then, with calculation of the scaled-dot product, we obtain the local temporal attention $(a(x_i^{(T)}, M_i ))$.
 
The middle block in figure \ref{img:threeattention} show the modified transformer block applicated local temporal attention.
And it replaces the transformer block in encoder in figure \ref{img:architecture}.
With this change, we confirm that despite the local window masking is simple, local temporal attention is effective for fMRI data.

\subsection{ROI-rank Based Masking}
\label{sec:preprosed method 5}

\begin{figure}[!t]
\centering
\includegraphics[width=1\textwidth]{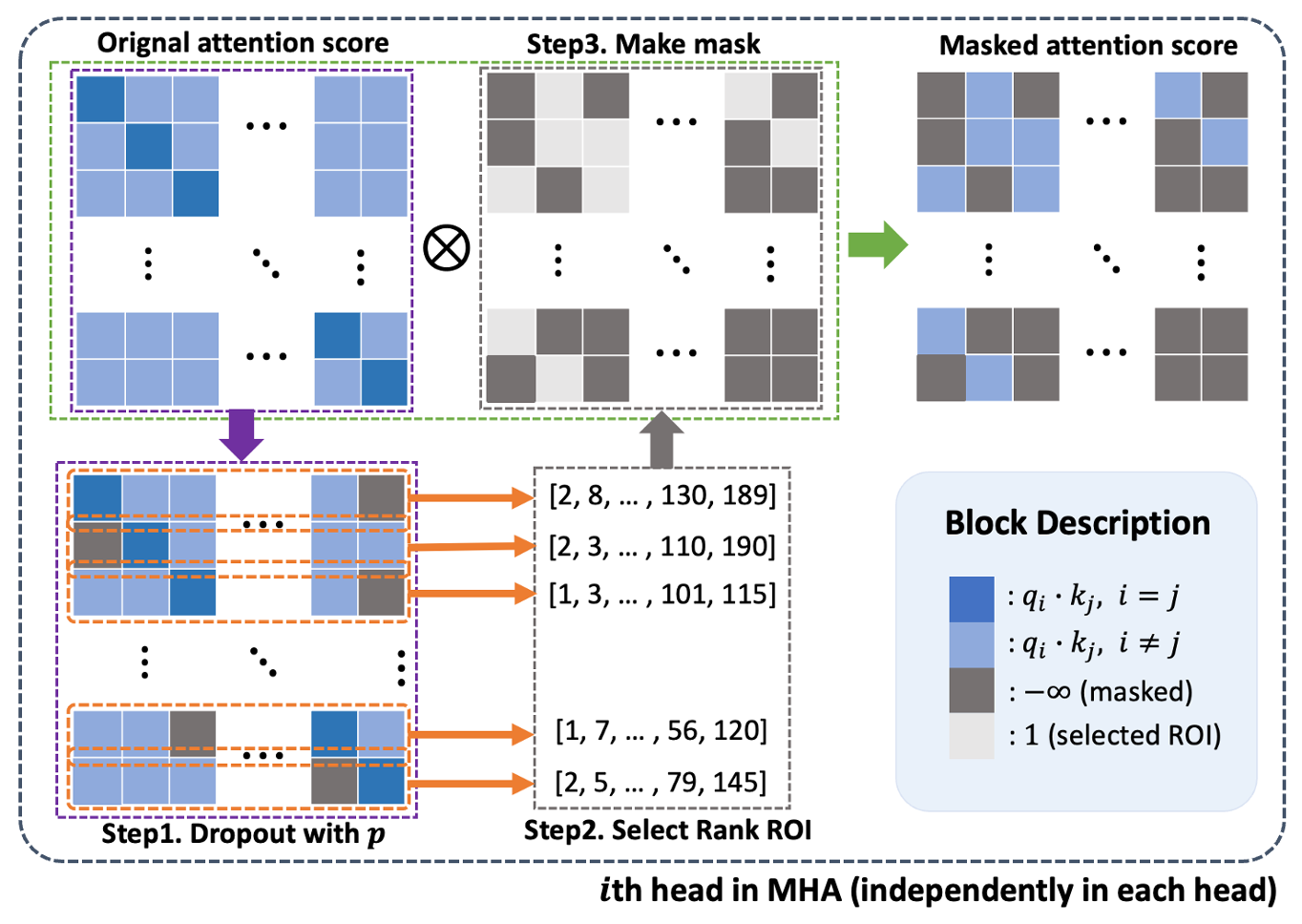}
\caption{The overall process of the ROI-rank based masking. It contains 3 steps. All steps are processed independently in each query axis and each head. At first, the attention score matrix is applied to the dropout layer. Next, step 2 lists the score values descending way and selects ranked ROIs in top-k. Finally, in step 3, make up the mask that only selected ROIs correspond to ‘1’ else ‘0’.}
\label{img:roirankmasking}
\end{figure}

Several studies on ADHD have shown that ADHD patients has notable activation in few brain regions compared with HC.
Therefore, we suppose that distinguishing some of important ROIs that hardly related with ADHD will have more accurate diagnostic performance. 
To confirm the clue to this hypothesis, we propose ROI-rank based masking, a variant of self-attention method for spatial perspective feature $(X^{(S)})$ in decoder.

ROI-rank based masking creates an attention mask to select and focus on important ROIs based on the attention score. 
In the attention mechanism, the attention score is obtained as the inner product of the query and the key $(s_{ij}=q_{i} \dot k_{j}^{T})$. 
And it indicates how importantly the value $(v_j)$ is to be considered. 
The attention score $(s_{ij})$ of self-attention for spatial perspective feature $(X^{(S)})$ indicates how closely a $i$th ROI is related to the $j$ th ROI.
By utilizing this, ROI-rank based masking selects most closely related ROIs with respect to the ith ROI, so that it can focus on relationships between more relatively important ROIs.

The detail of ROI-rank based masking is as shown in figure \ref{img:roirankmasking}. 
ROI-rank based masking is performed in 4 steps, and all ROI-rank based masking is applied head-independently to learn more diverse ROI relationships.

The first step (purple box) is the dropout to the score matrix to randomly select various ROI relationships regardless of the score.
Next, in the second step (orange box), with the largest attention score values from the score matrix, number of ranked ROI represented to index.
Currently, the selected ROIs differ depending on the target ROI.
Therefore, the rank ROI index list shape $(S \times R_k)$ is obtained corresponding to the product of the total number of ROIs $(S)$ and the number of Rank ROIs $(R_k)$.
In the third step, based on the rank ROI index list, the rank ROI mask is constructed.
The rank ROI mask is represented with two types of value (1 for selected Rank ROI index and $-\infty$ for otherwise).
And finally, in last step, the rank ROI mask is multiplied with original score matrix to get the masked attention score.

Through experiments, ROI-rank based masking was able to significantly improve the diagnostic performance of ADHD by only the relationships between a small number of ROIs.
Especially, in the experiment with datasets that preprocessed by various ROI templates, all showed improved diagnostic performance than previous.
Therefore, it can be a clue supporting the hypothesis that proposed earlier.

\section{Experimental Setting and Results}
\label{sec:experimental setting and results}
In this section, we describe the detail of experimental setting and the results.
To evaluate the our proposed transformer model for diagnosis ADHD, we experiment with several type of transformer.
And we further evaluate the model with different template of ADHD.
As a result, our proposed transformer model proved new aspect to diagnose ADHD.

\subsection{Dataset and Preprocessing for Experiment}
\label{sec:experimental setting and results 1}

The dataset used for the experiment was ADHD-200 \cite{RN17}, which consisted of rs-fMRI data provided by eight different sites (NYU : New York University child study center, Peking : Peking University, OHSU : Oregon Health Sciences University, KKI : Kennedy Krieger Institute, NI : NeuroIMAGE, BHBU : Bradley/ Brown University, Pitt : University of Pittsburgh, WUSTL : Washington University at Saint Louis).
They have pairs of rs-fMRI data and phenotypic information for 973 individuals.
Specifically, it is divided into a training dataset consisting of 776 individuals and an test dataset consisting of 197 individuals.
We selected data from 969 individuals based on the experimental criteria.
The detail composition of our dataset is referred in Table \ref{tab:traindataset} and \ref{tab:testdataset}.

Next, the rs-fMRI data are preprocessed with the Athena pipeline\footnote{Neuro Bureau Athena pipeline. Accessed: Mar. 7, 2025. [Online.]\\ Available: \underline{https://www.nitrc.org/plugins/mwiki/index.php/neurobureau:AthenaPipeline}}.
The Athena pipeline was performed on a site-by-site, and the specific preprocessing order is as follows.
At first step, it eliminates the noise due to the location and movement of the measurement device and individual through sling timing correction and motion correction.
The second step normalizes the brain size to the size of the s-MRI and then transform to the MNI152 space for standardization. 
The third step restricts the frequency band to match the resting state signal through denoising and applying a bandpass filter $(0.009 \mathrm{Hz} < f < 0.08 \mathrm{Hz})$.
Finally, spatial smoothing using a 6 mm FWHM Gaussian filter is performed to lose the variance of regional features.

Next, since the rs-fMRI data is too large to be used directly for learning, we reduced the dimension using a ROI-based template.
At this time, we used the structural-based templates AAL \cite{tzourio2002automated}, EZ \cite{eickhoff2005new}, HO \cite{goldstein2007hypothalamic}, and TT \cite{lancaster2000automated} and the functional-based template cc200 \cite{craddock2012whole}.
And the main ROI template in our experiment was cc200 because it empirically shows higher ADHD diagnostic performance than other templates.

Finally, the composition of the dataset for model training and evaluation followed the ADHD-200 competition.
Specifically, the train data (768 people) provided by all sites were used as the train set, and the test data (171 people) provided by all sites were used as the test set.
In addition, data corresponding to 20\% of the train set was randomly divided into a validation set.

\begin{landscape}
\begin{table*}
\setlength{\tabcolsep}{9pt} 
\renewcommand{\arraystretch}{1.3} 
\begin{center}
\begin{tabular}{| c | c | c | c | c | c | c | c | c | c |}
\hline
\multirow{2}{*}{Info} & \multirow{2}{*}{Type} & \multicolumn{7}{c|}{Template} & \multirow{2}{*}{Total} \\
\cline{3-9}
 & & KKI  & NI & NYU & OHSU & Peking & Pitt & WUSTL & \\ 
\hline
 HC & 0 & 61 & 23 & 98 & 42 & 116 & 89 & 59 & 488 \\ 
\cline{1-10}
 \multirow{3}{*}{ADHD}& 1 & 16 & 18 & 73 & 23 & 29 & 0 & 0 & 159 \\ 
\cline{2-10}
 & 2 & 1 & 6 & 2 & 2 & 0 & 0 & 0 & 11 \\ 
\cline{2-10}
 & 3 & 5 & 1 & 43 & 12 & 49 & 0 & 0 & 110 \\ 
\hline
 Total &  & 83 & 48 & 216 & 79 & 194 & 89 & 59 & 768 \\ 
\hhline{|=|=|=|=|=|=|=|=|=|=|}
\multirow{2}{*}{Gender} & F(0) & 37 & 17 & 78 & 35 & 52 & 43 & 27 & 289 \\ 
\cline{2-10}
 & M(1) & 46 & 31 & 137 & 44 & 142 & 46 & 32 & 478 \\ 
\hline
\multirow{2}{*}{Age} & mean & 10.24 & 16.99 & 11.67 & 8.83 & 11.98 & 15.11 & 11.59 & - \\ 
\cline{2-10}
 & std & 1.35 & 2.74 & 2.92 & 1.13 & 1.86 & 2.9 & 3.88 & - \\ 
\hline
\multirow{5}{*}{Handedness} & L(0) & 7 & 5 & 64 & 0 & 3 & 4 & 0 & 83 \\ 
\cline{2-10}
 & R(0) & 75 & 42 & 146 & 79 & 191 & 85 & 59 & 677 \\ 
\cline{2-10}
 & ambidextrous(2) & 1 & 0 & 0 & 0 & 0 & 0 & 0 & 1 \\ 
\cline{2-10}
 & unknown(3) & 0 & 1 & 0 & 0 & 0 & 0 & 0 & 1 \\ 
\cline{2-10}
 & None & 0 & 0 & 9 & 0 & 0 & 0 & 0 & 9 \\ 
\hline
\multirow{3}{*}{Full4 IQ} & mean & 110.01 & None & 108.3 & 113.76 & 113.02 & 109.8 & 116.05 & - \\ 
\cline{2-10}
 & std & 11.94 & None & 14.33 & 14.02 & 14.66 & 11.53 & 14.27 & - \\ 
\cline{2-10}
 & error value & 0 & 0 & 10(-999) & 0 & 1(-999) & 0 & 0 & 11 \\ 
\hline
\end{tabular}
\caption{Train dataset’s phenotypic information in each site.}
\label{tab:traindataset}
\end{center}
\end{table*}
\end{landscape}

\begin{landscape}
\begin{table*}
\setlength{\tabcolsep}{9pt} 
\renewcommand{\arraystretch}{1.3} 
\begin{center}
\begin{tabular}{| c | c | c | c | c | c | c | c | c |}
\hline
\multirow{2}{*}{Info} & \multirow{2}{*}{Type} & \multicolumn{6}{c|}{Template} & \multirow{2}{*}{Total} \\
\cline{3-8}
 & & KKI  & NI & NYU & OHSU & Peking & Pitt & \\
\hline
 HC & 0 & 8 & 14 & 12 & 28 & 27 & 5 & 94 \\
\cline{1-9}
 \multirow{3}{*}{ADHD}& 1 & 3 & 11 & 22 & 4 & 9 & 0 & 49 \\
\cline{2-9}
 & 2 & 0 & 0 & 0 & 1 & 1 & 0 & 2 \\
\cline{2-9}
 & 3 & 0 & 0 & 7 & 1 & 14 & 4 & 26 \\
\hline
 Total &  & 11 & 25 & 41 & 34 & 51 & 9 & 171 \\
\hhline{|=|=|=|=|=|=|=|=|=|}
\multirow{2}{*}{Gender} & F(0) & 1 & 13 & 13 & 17 & 19 & 2 & 65\\
\cline{2-9}
 & M(1) & 10 & 12 & 28 & 17 & 32 & 7 & 106 \\
\hline
\multirow{2}{*}{Age} & mean & 10.06 & 18.87 & 10.73 & 9.71 & 10.64 & 14.82 & - \\
\cline{2-9}
 & std & 1.34 & 3.28 & 2.72 & 1.33 & 1.96 & 1.12 & - \\
\hline
\multirow{5}{*}{Handedness} & L(0) & 2 & 3 & 14 & 0 & 1 & 0 & 20 \\
\cline{2-9}
 & R(0) & 9 & 20 & 26 & 34 & 50 & 9 & 148 \\ 
\cline{2-9}
 & ambidextrous(2) & 0 & 0 & 0 & 0 & 0 & 0 & 0 \\ 
\cline{2-9}
 & unknown(3) & 0 & 0 & 0 & 0 & 0 & 0 & 0 \\ 
\cline{2-9}
 & None & 0 & 0 & 1 & 0 & 0 & 0 & 1 \\ 
\hline
\multirow{3}{*}{Full4 IQ} & mean & 113.81 & None & 106.38 & 113.85 & 112.92 & 107 & - \\ 
\cline{2-9}
 & std & 7.92 & None & 14.32 & 12.67 & 13.4 & 13.13 & - \\
\cline{2-9}
 & error value & 0 & 0 & 0 & 0 & 0 & 0 & - \\
\hline
\end{tabular}
\caption{Test dataset’s phenotypic information in each site.}
\label{tab:testdataset}
\end{center}
\end{table*}
\end{landscape}

\begin{table}
\setlength{\tabcolsep}{9pt} 
\renewcommand{\arraystretch}{1.0} 
\begin{center}
\begin{tabular}{| c | c |}
\hline
Parameter name & Setting \\
\hhline{|=|=|}
Number of transfomer block & 2 \\
\hline
embedding hidden size $(d_{model})$ & 256 \\
\hline
Attention hidden size $(d_{a})$ & 256 \\
\hline
Feed-Forward hidden size $(d_{ff})$ & 1024 \\
\hline
Number of head $(h)$ & 8 \\
\hline
Rate of dropout $(p_{drop})$ & 0.1 \\
\hline
Transformer initialization & Truncated Normal $(0, \sqrt{ 1 \over d_{model}})$ \\
\hline
Activation layer & Gelu \\
\hline
Classification layer hidden sizes & [256, 10, 1] \\
\hline
Classification layer initialization & [He, He, Glorot] \\
\hline
\end{tabular}
\caption{Detail of model parameter setting.}
\label{tab:modelparameter}
\end{center}
\end{table}

\begin{table}
\setlength{\tabcolsep}{35pt}
\renewcommand{\arraystretch}{1.0}
\begin{center}
\begin{tabular}{| c | c |}
\hline
Parameter name & Setting \\
\hhline{|=|=|}
fMRI signal segment length & 60 \\
\hline
Training epoch & 30 \\
\hline
Optimizer & Adam \\
\hline
Learning rate & 1e-5 \\
\hline
Batch size & 128 \\
\hline
\end{tabular}
\caption{Detail of training parameter setting.}
\label{tab:trainingparameter}
\end{center}
\end{table}

\begin{table}
\begin{center}
\renewcommand{\arraystretch}{1.0}
\begin{tabular}{| c | c | c | c | c |}
\hline
Model & Architecture & Features & Site & ACC \\
\hhline{|=|=|=|=|=|}
\multicolumn{5}{|c|}{Previous proposed model} \\
\hline
3D CNN & 3D CNN & \begin{tabular}[c]{@{}c@{}}fALFF\\ GM density\end{tabular} & PK, KKI, NYU & 69.15 \\
\hline
4D CNN & 4D CNN & raw fMRI & \begin{tabular}[c]{@{}c@{}}PK, KKI, NI\\ NYU, OHSU\end{tabular} & 71.3 \\
\hline
EM-MI & \begin{tabular}[c]{@{}c@{}}4D CNN\\ short time\end{tabular} & AAL 116 & \begin{tabular}[c]{@{}c@{}}PK, KKI, NI\\ NYU, OHSU\end{tabular} & 70.4 \\
\hline
FC-HAT & \begin{tabular}[c]{@{}c@{}}Hypergraph\\ attention\end{tabular} & AAL 90 & All & 69.2 \\
\hline
STAAE & \begin{tabular}[c]{@{}c@{}}Auto-encoder\\ self-attention\end{tabular} & MNI 152 & \begin{tabular}[c]{@{}c@{}}PK, KKI, NI\\ NYU, OHSU\end{tabular} & 72.5 \\
\hline
Trans3D & \begin{tabular}[c]{@{}c@{}}Encoder\\ transformer\end{tabular} & raw fMRI & \begin{tabular}[c]{@{}c@{}}PK, KKI, NI\\ NYU, OHSU\end{tabular} & 74.5 \\
\hline
STCAL & \begin{tabular}[c]{@{}c@{}}Vision\\ transformer\end{tabular} & CC 200 & All & 74.3 \\
\hline
\multicolumn{5}{|c|}{Our proposed model} \\
\hline
\begin{tabular}[c]{@{}c@{}}Baseline\\ transformer\end{tabular} & \begin{tabular}[c]{@{}c@{}}Enc-dec\\ transformer\end{tabular} & CC 200 & All & 70.76 \\
\hline
\begin{tabular}[c]{@{}c@{}}Enhanced\\ Transformer\end{tabular} & \begin{tabular}[c]{@{}c@{}}Enc-dec\\ transformer\end{tabular} & \begin{tabular}[c]{@{}c@{}}CC 200\\ (60 Rank)\end{tabular} & All & \textbf{77.8} \\
\hline
\end{tabular}
\caption{ADHD diagnosis accuracy comparison with previous model.}
\label{tab:modelcomparison}
\end{center}
\end{table}

\subsection{Detail of Model Setting and Evaluation}
\label{sec:experimental setting and results 2}

The model configured for the experiment is as follows.
First, two different linear layers were used for embedding to each perspective features.
Then, two transformer blocks were used in each encoder and decoder part.
The exception is that only the transformer block’s attention method at the end of the decoder was replaced to co-attention.
Then, the output of the decoder was propagated to the classification layers.
At this time, the classification layers was consisted with three fully connected neural networks.
And all activation functions in the model used GeLU.
The detail of model parameter setting is referred in Table \ref{tab:modelparameter}.

Next, the training parameters are as in the Table \ref{tab:trainingparameter}.
Specifically, to supplement the insufficient fMRI data, we randomly cut the time features from the original fMRI data into sizes of 60 and performed data augmentation during model training.
And for a fair evaluation, both the validation and test datasets were cut into the same size of 60 as the center part of the rs-fMRI signal.
And for model evaluation, accuracy (ACC), specificity (SPE), sensitivity (SEN), and area under the receiver operation characteristic curve (AUC), which are the main indices for evaluating diagnostic models, were selected.
All experiments are evaluated with selected model that achieved the best performance through the validation set.

\subsection{Comparison with Other Previous Models}
\label{sec:experimental setting and results 3}

We compared the ADHD diagnostic ability of our proposed transformer model with the performance of several existing fMRI models.
At this time, the comparison models were selected based on the CNN and attention for ADHD diagnosis problems.
A brief description of the selected comparison models is as follows.

\begin{enumerate}{}{}
    \item{3D CNN \cite{RN1}: 3D CNN is a model that effectively learns the spatial features of two different MRI data through a deep learning model composed of 3D convolution layers.
    By using fMRI and structural MRI (sMRI) scanned from the identical subject, it showed that the two features have a complementary relationship in extracting information from brain.
    Through this, it achieved an accuracy that was 3-6\% higher than that of existing traditional machine learning models.
    }
    \item{4D CNN \cite{RN8}: It compensates the spatial and temporal features learning independently with parallelizes three models.
    It is composed with 4D channel convolution layer and 4D spatial convolution layer.
    Through this, they showed that learning the features of fMRI at granularity is important for ADHD diagnosis.}
    \item{EM-MI \cite{dou2020adhd}: This paper proposes the EM-MI algorithm, a method to distinguish key-frames from rs-fMRI data for rapid diagnosis of ADHD.
    They used the model that is a C4N network based on 4D CNN.
    Through experiments, it is shown that ADHD can be diagnosed quickly and effectively with only a small number of frames, and the importance of the temporal features of fMRI in ADHD diagnosis is demonstrated.}
    \item{FC-HAT \cite{ji2022fc}: FC-HAT is a model based on hypergraph network for high-dimensional functional brain network analysis.
    Hypergraph network has the advantage of enabling higher-dimensional information extraction than existing graph networks.
    In addition, the result showed that the model can effectively organize hypergraph clusters by combining with the attention method.
    As a result, they improved the diagnosis performance than existing network-based models.
    }
    \item{STAAE \cite{qiang2022novel}: STAAE is an auto-encoder model for spatial-temporal feature learning utilizing self-attention proposed by transformer. 
    With the unsupervised learning, self-attention can learn a wide range of temporal features from fMRI data. 
    In addition, it achieves more effective diagnostic ability than existing ICA-based RSN features by using the encoder features of STAAE pre-trained with rs-fMRI.}
    \item{Trans3D \cite{qin2022ensemble}: Trans 3D is a variant model of vision transformer (ViT), the transformer model for image. 
    It refined to learn fMRI data deal with as like a 3D image. 
    To do this, it embeds fMRI data into 3D small-patch spatial information using a 3D convolution layer. 
    In addition, key phenotypic data obtained by random forest is used together for ADHD diagnosis. 
    It achieved considerably high performance.}
    \item{STCAL \cite{liu2023spatial}: STCAL is a variant model of transformer. 
    This model consists of two self-attention modules for independent learning of spatial and temporal features of fMRI and a guide co-attention module with transition block to switch the attention score between two different features. 
    Through this, it showed the importance of utilizing two features of fMRI together.}
\end{enumerate}

ADHD diagnosis models using fMRI are difficult to compare under the same conditions because they use different data for training and evaluation.
Therefore, we have written the experiment setting of comparison models in Table \ref{tab:modelcomparison}.
Specifically, the ADHD-200, data preprocessing method, and accuracy calculation method for each model.

In Table \ref{tab:modelcomparison}, our proposed encoder-decoder based Transformer for diagnosis ADHD shows a comparable performance to that of existing CNN-based models (3D CNN, 4D CNN) with (70.76ACC 69.15SPE 72.73SEN 73.96 AUC).
While existing CNN-based models focus on training the local temporal feature, the transformer structurally trains the temporal features globally.
Nevertheless, it is positive in terms of model adequacy that it records high ADHD diagnosis performance.
Furthermore, with our proposed modifications in transformer, we achieve 7\% more higher performance (77.78ACC 78.72SPE 76.62SEN 79.37AUC) than before.
This result not only surpasses the performance of existing CNN-based models, but also surpasses the performance of the transformer-based ADHD diagnosis model.

The reason why our modified transformer can achieve such performance is because it refocuses on training of local temporal features of fMRI data through local temporal attention and CNN-based embedding.
And it can selectively focus on the relationships between some of small number of ROIs with Rank-based masking.
Therefore, as proven in the previous models (e.g., STCAL, Trans3D), configuring the embedding method and the attention method for local temporal feature extraction of fMRI are effective in the transformer model.
Also, it is important to distinguish ROIs which are more associated with ADHD diagnosis.

In summary, the temporal and spatial feature learning of fMRI data can be effectively performed with only three simple modifications of the transformer-based ADHD diagnosis.
In particular, the three modifications surpassed existing models’ result.
For more specific experiment and analyses for the three methods can be found in the following subsections.

\subsection{Comparison with Variant Transformer Structure}
\label{sec:experimental setting and results 4}

\begin{figure}[!t]
\centering
\includegraphics[width=0.5\textwidth]{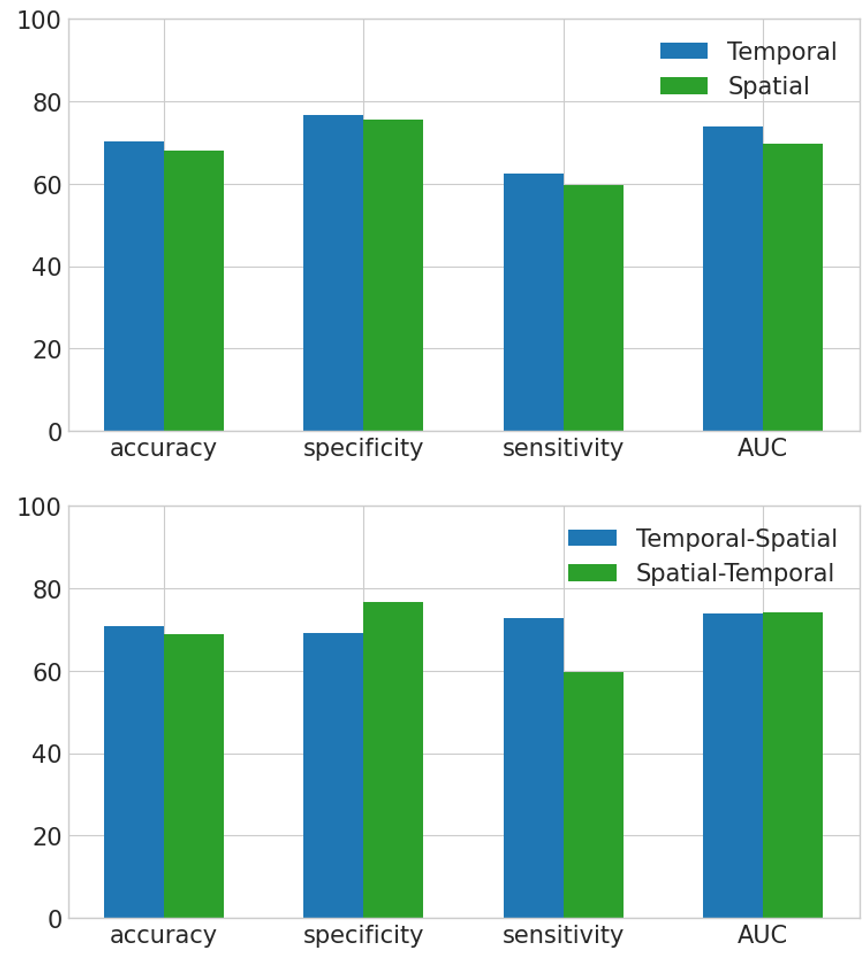}
\caption{The model performance comparison between different perspective input features in transformer for diagnosis ADHD. The upper figure is the result comparison with encoder-only transformers. The lower figure is the result comparison with encoder-decoder transformers.}
\label{img:transformerbarplot}
\end{figure}

We compared the variant type of transformer structure when using fMRI data.
The transformer model is divided into two types: the encoder-only and the encoder-decoder, depending on the fMRI input method.
Specifically, in figure \ref{img:architecture}, the encoder-only model uses only the encoder part of the transformer.
It means that uses one perspective fMRI feature.
On the other hand, described in section \ref{sec:preprosed method 2}, the encoder-decoder model use both perspective fMRI features together.
In this case, it is possible to use temporal perspective feature as input to the encoder and spatial perspective features as input to the decoder, or vice versa.

The figure \ref{img:transformerbarplot} show the comparison results in each structure.
In upper figure in figure \ref{img:transformerbarplot}, temporal perspective feature result (70.18ACC 76.60SPE 62.34SEN 73.92AUC) outperforms spatial feature result (68.42ACC  75.53SPE 59.74SEN 69.74AUC) as input data.
It means that the attention model is more suitable for learning temporal perspective features.
However, it is only a relative result. spatial perspective features are also essential for ADHD diagnosis.
Because the encoder-decoder model, which can learn two fMRI features together, outperforms the encoder-only model that learns only one feature.
Specifically, when the input to the encoder is temporal perspective feature and the input to the decoder is spatial perspective feature, the performance is (70.76ACC 69.15SPE 72.73SEN 73.96 AUC), which is higher than all existing encoder-only models.
Also, when the input data is used conversely, this performance is better than used only spatial perspective features in all evaluations (69.01 ACC 76.60 SPE 59.74 SEN 74.28 AUC).

In summary, temporal perspective feature is crucial rule in diagnosis ADHD. 
However, training the model without spatial perspective feature is not enough to extract entire information from rs-fRMI.
Therefore, it is essential to complement each perspective features.
For that reason, we improved this transformer more suitable with rs-fMRI to diagnose ADHD.

\subsection{Searching the CNN-based Embedding Method}
\label{sec:experimental setting and results 5}

\begin{table}
\setlength{\tabcolsep}{11pt}
\renewcommand{\arraystretch}{1.2}
\begin{center}
\begin{tabular}{| c | c | c | c | c |}
\hline
\multirow{2}{*}{\# of CNN blocks} & \multicolumn{4}{c|}{Evaluation results} \\
\cline{2-5}
 & ACC & SPE & SEN & AUC\\
\hline
0 (Linear embedding) & 70.76 & 69.15 & 72.73 & 73.96 \\
\hhline{|=|=|=|=|=|}
2 & 66.08 & 78.72 & 50.65 & 74.97 \\
\hline
3 & \textbf{71.34} & \textbf{72.34} & 70.13 & \textbf{76.25} \\
\hline
4 & 64.91 & 71.28 & 57.14 & 74.06 \\
\hline
5 & 61.40 & 55.32 & 68.83 & 70.13 \\
\hhline{|=|=|=|=|=|}
3 (Enhanced) & \textbf{73.10} & \textbf{73.40} & \textbf{72.73} & \textbf{77.63} \\
\hline
\end{tabular}
\caption{The result comparison with different number of CNN block in CNN-based embedding.}
\label{tab:numberofcnnblock}
\end{center}
\end{table}

we conducted a structural search for a proper CNN-based embedding module to replace the linear embedding of transformer. 
To this end, we experimented with ADHD diagnosis performance according to the number of the CNN block consisting of convolution layers and pooling layers.
The CNN-based embedding module was designed based on the CNN structure of FCNet \cite{riaz2017fcnet}, and the detail can be seen in figure \ref{img:cnnembedding}.
For the experiment, we used encoder-decoder (temporal-spatial) transformer.
And the embedding layer of the decoder, which embeds spatial perspective features, was changed from the existing linear embedding layer to a CNN-based embedding module.

Table \ref{tab:numberofcnnblock} shows the performance evaluation results according to the number of CNN blocks increased from 2 to 5.
In this results, performance changes significantly as the number of CNN blocks increases.
Specifically, the model with the highest performance has a structure that uses 3 CNN blocks, which has higher performance (71.34ACC 72.34SPE 70.13SEN 76.25AUC) than the existing linear embedding
However, on the other hand, when 5 CNN blocks are used, the performance is far lower than the linear embedding (61.40ACC 55.32SPE 68.83SEN 70.13AUC).
This is because attention learning is sensitive to the embedding method of the transformer model.
Therefore, performance can be improved only when CNN blocks are used appropriately.

Furthermore, we added a convolution layer without pooling only to the last CNN block, like the FCNet, to further enhance the performance of the three CNN block embeddings.
The specific structure is as shown right block in figure \ref{img:cnnembedding} and the channel size in this block is [32, 64, 256, 256].
As a result, we were able to obtain a 2\% performance improvement (73.10 ACC 73.40 SPE 72.73 SEN 77.63 AUC) compared to the case where three CNN blocks were used.
In other words, it was shown that it is important to train the channel features in deep layers because the complexity of the features contained in the fMRI signal is high.

\subsection{Analysis with Local Temporal Attention}
\label{sec:experimental setting and results 6}

\begin{figure}[!t]
\centering
\includegraphics[width=0.9\textwidth]{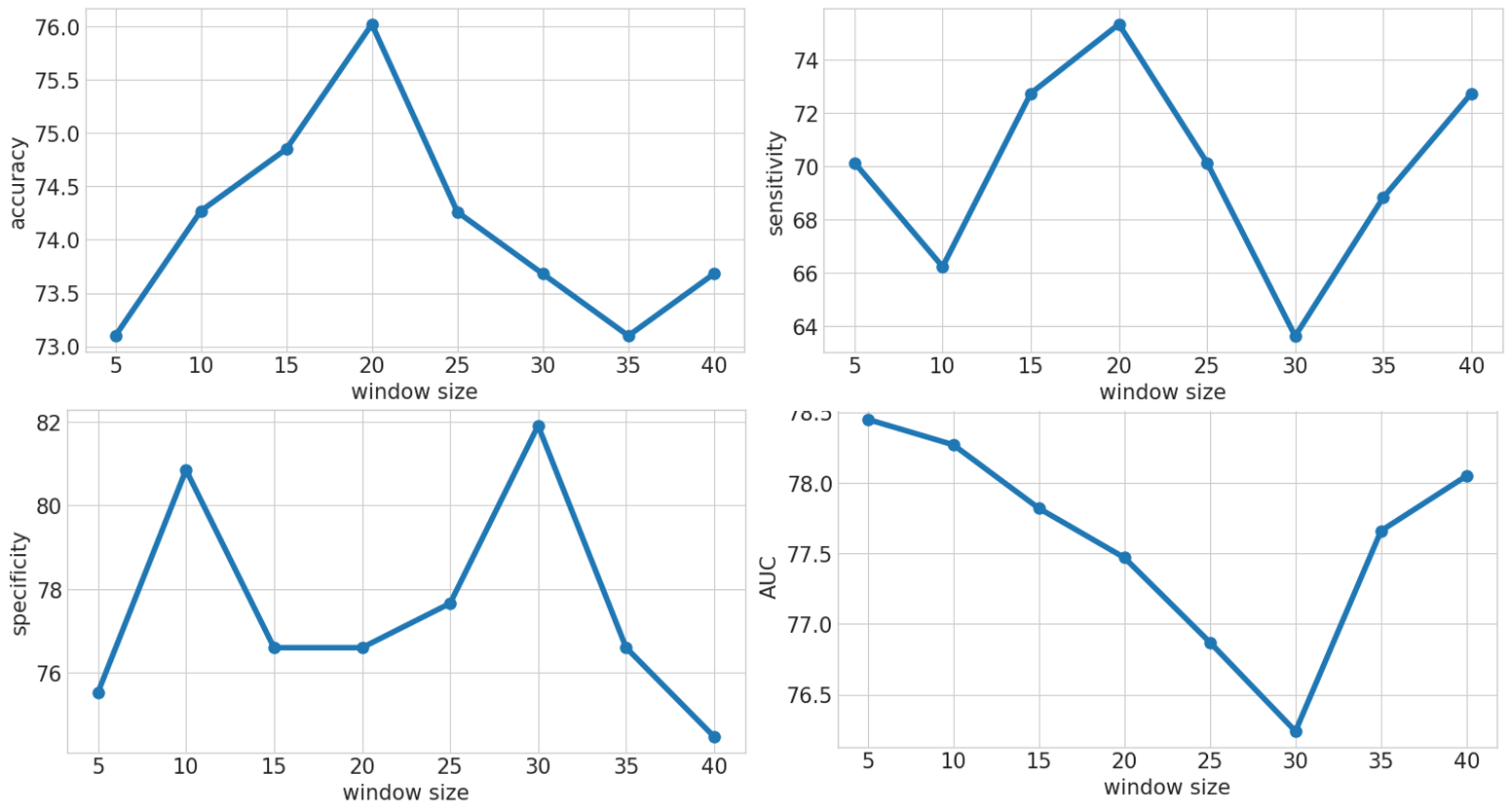}
\caption{ADHD diagnosis performance plots according to window size changes in local temporal attention.}
\label{img:windowresult}
\end{figure}

\begin{table}
\setlength{\tabcolsep}{11pt}
\renewcommand{\arraystretch}{1.2}
\begin{center}
\begin{tabular}{| c | c | c | c | c |}
\hline
\multirow{2}{*}{Position in encoder} & \multicolumn{4}{c|}{Evaluation results} \\
\cline{2-5}
 & ACC & SPE & SEN & AUC\\
\hline
no use & 73.10 & 73.40 & 72.73 & \textbf{77.63} \\
\hline
1 & \textbf{76.02} & \textbf{76.60} & \textbf{75.32} & 77.47 \\
\hline
2 & 71.93 & 72.34 & 71.43 & 76.03 \\
\hline
All & 74.27 & 77.66 & 70.13 & 76.87 \\
\hline
\end{tabular}
\caption{Local temporal attention performance comparison applicated in different position.}
\label{tab:positionlocaltemporalattention}
\end{center}
\end{table}

We analyses the ADHD diagnosis performance with local temporal attention.
In the experiment, the applied location of local temporal attention in transformer is the encoder part of corresponding to temporal perspective feature.

Figure \ref{img:windowresult} shows the results of the diagnostic performances with different window size.
In this case, local temporal attention was applied only to the first layer of the encoder.
These results showed that the diagnostic performance is always improved regardless of the window size than before.
Specifically, when the window size is 20, it shows the highest diagnostic performance (76.02 ACC 76.60 SPE 75.32 SEN 77.47 AUC).
This is 1-2\% higher performance than other case.
Therefore, it is important to identify the fMRI signal’s relationship within a specific length in diagnosing ADHD.

Next, Table \ref{tab:positionlocaltemporalattention} shows the diagnosis results according to the application position of local temporal  attention in transformer.
What is different from the previous results is that using local mask attention in all encoder blocks does not improve the performance.
Also, the performance when applied only to the second encoder block (71.93ACC 72.34SPE 71.43SEN 76.03AUC) is lower than before even though local mask attention was used.
This shows that for effective temporal feature learning, it is necessary to gradually increase the range from local features of the fMRI signal to global features as the layer deepens.

\subsection{Analysis with ROI-Rank Based Masking}
\label{sec:experimental setting and results 7}

\begin{figure}[!t]
\centering
\includegraphics[width=0.9\textwidth]{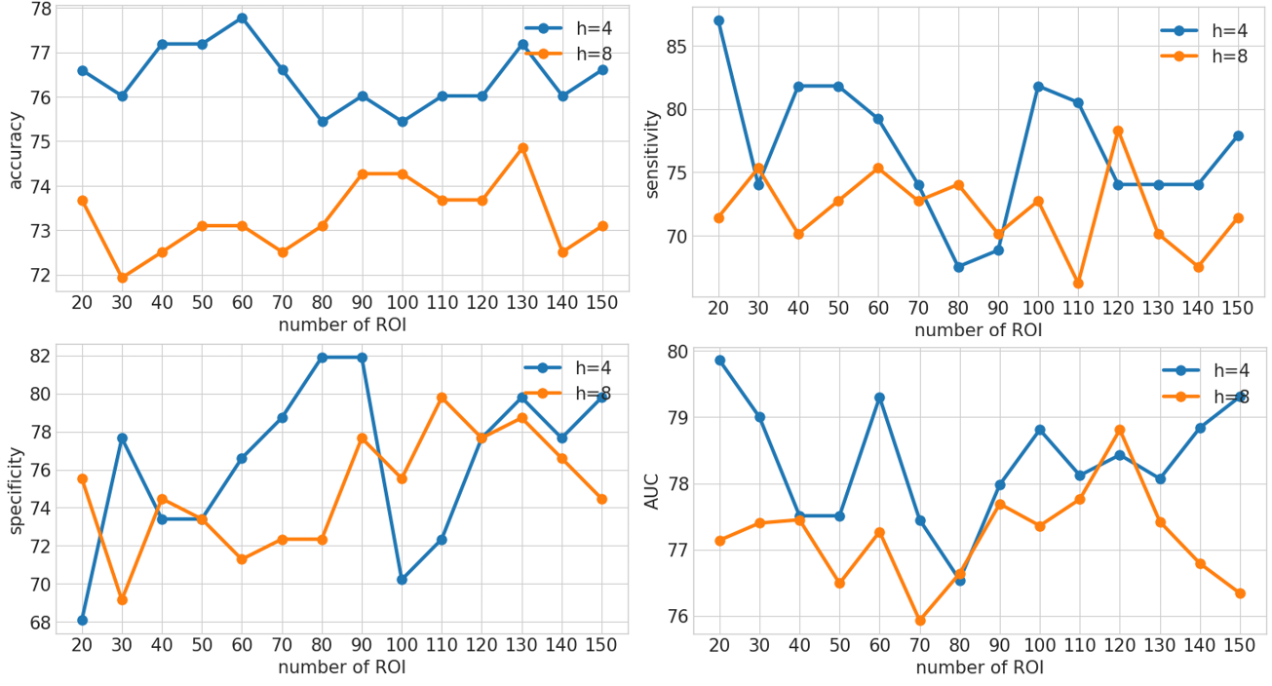}
\caption{ADHD diagnosis performance plots according to top-k of ROI selected in Rank-ROI based masking. Specifically, with 4 heads (blue line) shows more higher performance than 8 heads (orange line).}
\label{img:roirankresult}
\end{figure}

We analysis the ADHD diagnosis performance according to the number of ROIs used with Rank-based masking.
Figure \ref{img:roirankresult} shows the ADHD diagnosis performance according to the number of ROIs used.
And, the experiment was conducted twice by setting of the number of heads in the decoder to 4 and 8.

Specifically, using only the 60 top-scoring ROIs with four heads, we achieved the best performance for ADHD diagnosis (77.78 ACC 76.60 SPE 79.22 SEN 79.30 AUC).
It is significantly interesting result. 
Because, despite using fewer ROIs, it is about 2\% higher performance than the case of using 190 ROIs in the existing decoder (76.02 ACC 76.60 SPE 75.32 SEN 77.47 AUC).

These results suggest two implications.
First, it is not essential to consider all brain regions to diagnose ADHD.
In other words, distinguishing important ROIs for diagnosing ADHD can lead to a more accurate diagnosis.
Second, it is possible to select important ROIs in attention and learn their relationships with Rank-based masking.
In particular, the attention method is more suitable for training fMRI data because it learns by selecting important ROIs due to the mechanism structure of how it works.

Additionally,the experimental results show that the performance differs by approximately 4\% when the number of heads is 4 and 8.
This seems when attention head decreases in decoder, it much well-fit for simpler relationship between small amout of ROIs.
Therefore, by reducing the number of heads of the decoder, the relationship between important ROIs can be trained more clearly
Thereby, the ADHD diagnosis ability is highly improved.

\subsection{Performance Analysis with Different Template of ADHD}
\label{sec:experimental setting and results 8}

\begin{table}
\setlength{\tabcolsep}{9pt}
\renewcommand{\arraystretch}{1.2}
\begin{center}
\begin{tabular}{| c | c | c | c | c |}
\hline
\multirow{2}{*}{Template (\# of Rank ROI)} & \multicolumn{4}{c|}{Evaluation result} \\
\cline{2-5}
 & ACC & SPE & SEN & AUC\\
\hline
\multicolumn{5}{|c|}{ROI-rank masking not used} \\
\hline
CC 200 & \textbf{76.02} & \textbf{76.06} & 75.32 & 77.47 \\
\hline
AAL 116 & 74.85 & 75.53 & 74.02 & \textbf{77.94} \\
\hline
EZ 116 & 75.43 & 74.47 & 76.62 & 77.27 \\
\hline
HO 111 & 72.51 & 62.77 & \textbf{84.42} & 77.12 \\
\hline
TT 97 & 70.76 & 63.83 & 79.22 & 76.54 \\
\hline
\multicolumn{5}{|c|}{ROI-rank masking used} \\
\hline
CC 200 (60) & \textbf{77.78} & \textbf{76.60} & 79.22 & \textbf{79.30} \\
\hline
AAL 116 (70) & 75.44 & 71.28 & 80.52 & 77.45 \\
\hline
EZ 116 (60) & 76.02 & 73.40 & 79.22 & 77.43 \\
\hline
HO 111 (80) & 73.68 & 61.70 & \textbf{88.31} & 77.60 \\
\hline
TT 97 (30) & 74.85 & 69.15 & 81.82 & 78.21 \\
\hline
\end{tabular}
\caption{ADHD diagnosis performance comparison with different ROI templates.}
\label{tab:differenttemplates}
\end{center}
\end{table}

we investigated the effects of different ROI-based template preprocessing methods in the transformer-based model on ADHD diagnosis. 
To this end, we conducted additional experiments with different ROI-based preprocessing templates from the ADHD200 dataset. 
The information on the preprocessing templates used and the total number of ROIs is as follows (AAL: 116, EZ: 116, HO: 111, TT: 97). 
In addition, As an extension of the previous section \ref{sec:experimental setting and results 7}, we intend to see how ROI selection with rank ROI masking affects ADHD diagnosis in other types of preprocessing templates.

Table. \ref{tab:differenttemplates} shows the results of the experiment. 
First, when comparing the performance of each ROI-based preprocessing template for ADHD diagnosis, the performance decreases in the following order (cc200 – EZ – AAL – HO – TT). 
This is the same as the order in which the number of ROIs distinguished by template decreases. 
Specifically, the performance difference between the highest performing template, CC200, and the lowest performing template, TT, is 6\%, and the difference in the number of ROIs is approximately twice. 
Based on this, we suggest that dividing more brain regions to train the transformer model is more helpful for ADHD diagnosis.

Another notable thing is that, for all templates, when selecting ROIs based on scores through Rank base ROI, diagnosis performance improved from 1\% to 4\%.
Interestingly, the TT template with the smallest ROI showed the largest improvement $(\sim 4\%$).
Therefore, finding and selecting ROIs can be an important component to effectively diagnosis ADHD.

\begin{figure}[!t]
\centering
\includegraphics[width=0.7\textwidth]{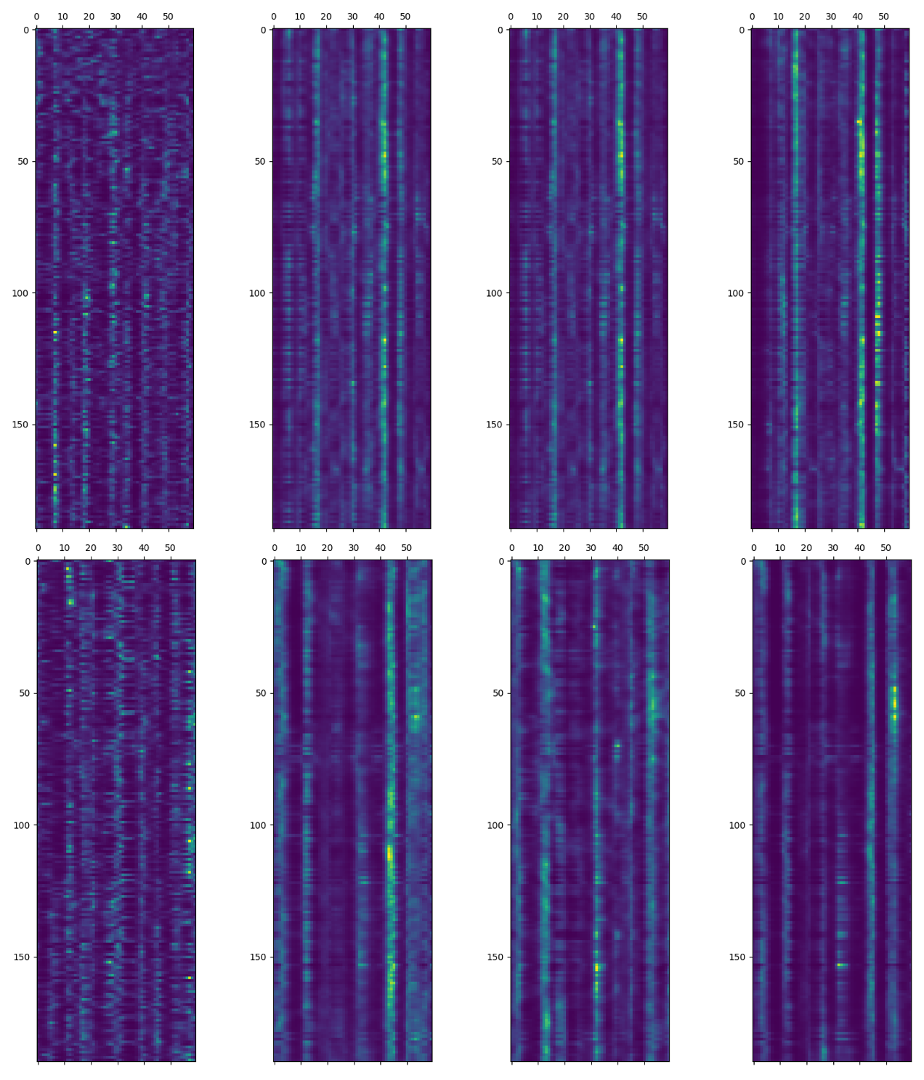}
\caption{The list of Co-attention score plot according to the transformer model’s change. From left to right, the results come from the trained models by sequentially adding the following methods (CNN-based Embedding - local attention - ROI-rank masking) based on the baseline transformer model. The subject’s phenotypic information are as follows. (upper) ID : 20001, site : KKI, HC (lower) ID : 2599965, site : Peking, ADHD.}
\label{img:scoreplot}
\end{figure}

\subsection{Co-attention Score Matrices Analysis}
\label{sec:experimental setting and results 9}

In order to figure out the specific learning tendency according to the proposed methods added, we plot the several co-attention scores that trained relation between spatiotemporal features in rs-fMRI.
Figure \ref{img:scoreplot} shows the list of the score plot according to the different transformer models.
The left figure is the result from the basic transformer model, and to the right, result come from the models with the proposed methods added in the following order (CNN-based Embedding - local attention - ROI-rank masking).

As the methods added (in figure \ref{img:scoreplot} left to right), an important thing is that the co-attention score becomes more distinguishable how much important between temporal and spatial features.
The points to note here are as follows:
In the case of the basic transformer model (left plot in figure \ref{img:scoreplot}), the points where temporal feature is importantly considered are highly dispersed.
On the other hand, in the case of the model with CNN-based Embedding and local attention added, the focused temporal points are revealed much more clearly than before.
It means that these two proposed modules can lead to effective temporal feature learning.

Next, the use of ROI-rank masking (right plot in figure \ref{img:scoreplot}) shows that ROI selection can learn ROIs which are important for ADHD judgment more clearly.
Specifically, looking at the example of ADHD patients (lower-right plot in figure \ref{img:scoreplot}), certain sparse ROIs (ROI index: $50 \sim 60$) have higher scores than other ROIs.
whereas, in example of HC patients (upper-right plot in figure \ref{img:scoreplot}), the highly activated ROI's number is more than ADHD patients.
Therefore, determining ADHD by relying on a few ROI features implies that ROI selection is an important problem for ADHD diagnosis.

\section{Conclusion}
\label{sec:conclusion}

We propose a novel transformer architecture that confirmed the important time and regions at once within a whole brain for ADHD diagnosis.
Because of the encoder-decoder structure, it can fully attention on not only independent but also correlation of spatial-temporal features.
Furthermore, to enhance the three part of transformer, we more concentrated on to learn the critical biomarker of ADHD.
The one is CNN-based embedding method, it can hierarchically embed BOLD signal of fMRI without losing these temporal information.
Next, for temporal feature attention, the local temporal attention lead to attend within local neighbor signals.
Finally, for spatial feature attention, the ROI-rank masking focus on connectivity of ROIs that are highly related with the ADHD.
Especially, the ROI-rank masking provided the new clue that is essential to find out what is the important sparse network in brain.
As the result, we achieved the outperformed diagnosis performances in ADHD.
Therefore, with development of these deep learning model, we hope to discover uncertainty of the ADHD occurrence.

\section*{Acknowledgments}
This work is supported by the Basic Science Research Program through the National Research Foundation of Korea (NRF) funded by the Ministry of Education, Science and Technology (NRF2022R1F1A1064459). 


\end{document}